\documentclass[prl, twocolumn,showpacs,nofootinbibfloatfix,amsmath,amsfonts,amssymb]{revtex4-1}%
\usepackage{amsmath,amsfonts,amssymb,color}
\usepackage{amsthm}
\usepackage{leftidx}
\usepackage{graphicx}
\usepackage{xcolor}
\usepackage{dcolumn}
\usepackage{bm}
\usepackage{epstopdf}
\usepackage{epsfig}
\usepackage{mathdots} 

\begin{document}
	\title{Quantum repetition codes as building blocks of large period discrete time crystals}
	\author{Raditya Weda Bomantara}
	\email{Raditya.Bomantara@sydney.edu.au}
	\affiliation{%
		Centre for Engineered Quantum Systems, School of Physics, University of Sydney, Sydney, New South Wales 2006, Australia
	}
	\date{\today}
	
	
	\vspace{2cm}
	
\begin{abstract}
	Discrete time crystals (DTCs) are nonequilibrium phases of matter with exotic observable dynamics. Among their remarkable features is their response to a periodic drive at a fraction of its frequency. Current successful experiments are however only limited to realizing DTCs with period-doubling { and period-tripling} observable dynamics, forming only a very small subset of DTC phases. Creating larger periodic DTCs in the lab remains a longstanding challenge, yet it is necessary for developing the technological applications of DTCs, { e.g., as a quantum memory for highly-entangled qubits,} or exploring interesting features beyond subharmonic dynamics, { e.g., condensed matter phenomena in the time domain}. By highlighting the connection between DTCs and quantum error correction, we devise a general and realistic scheme for building DTCs exhibiting any large period observable dynamics, { which is observable even at sufficiently small system sizes}. Our proposal uses an array of spin-1/2 chains to simulate a repetition code at the hardware level, which has essential properties to realize robust observable dynamics. It is readily implemented with existing superconducting or trapped-ion quantum processors, making new families of DTCs experimentally accessible in the immediate future.  
\end{abstract}

\maketitle

\textit{Introduction.} Discrete time crystals (DTCs) were proposed in \cite{DTC1,DTC2,DTC3,DTC4,DTC5} to demonstrate instances of spontaneous time-translational symmetry breaking phenomena in time-periodic systems. Such phases of matter serve as the nonequilibrium counterpart of the (continuous) time crystals \cite{TC1,TC2}, which are prohibited to exist in local time-independent setting \cite{NG1,NG2,NG3}. The area of DTCs has rapidly evolved over the years and motivated various theoretical \cite{DTC6,DTC7,DTC8,DTC9,DTC10,DTC11,DTC12,DTC13,DTC14,DTC15,DTC15b,DTC16,DTC17,DTC18,DTC19,DTC20,DTC21,DTC22,DTC23,DTC24,DTC25,DTC26,DTC27,DTC29,DTC30,DTC31,DTC32,DTC33} and experimental \cite{DTCexp1,DTCexp2,DTCexp3,DTCexp4,DTCexp5,DTCexp6,DTCexp7,DTCexp8} studies. Moreover, their potential applications for quantum computation \cite{DTCqc}, quantum simulation \cite{DTCqs,DTCqs2}, and exploring condensed matter phenomena in the time domain \cite{DTCcm1,DTCcm2,DTCcm3,DTCcm5,DTCcm6,DTCcm7,DTCcm8} have been envisioned.

The main signature of DTCs is the existence of observables displaying persistent $nT$ periodicity in the thermodynamic limit, where $T$ is the driving period and $n\neq 1$ is a fixed integer insensitive to slight variation of system parameters \cite{DTC2}. { In previous theoretical studies, constructing an $nT$-period DTC with $n>2$ typically involves particles with large degrees of freedom, e.g., bosonic \cite{DTC1,DTC18} or $n$-dimensional clock states \cite{DTC22}. In Refs.~\cite{DTC15,DTC15b}, it was demonstrated that a disorder-free continuously driven spin-1/2 system with Ising-like interaction can in fact support some larger-period prethermal DTCs, which are however only observable at sufficiently large system sizes, e.g. $\gtrapprox 100$ particles \cite{DTC15b},} and under a rather specific choice of initial states.
	
At present, current succesful experimental demonstrations of DTCs \cite{DTCexp1,DTCexp3,DTCexp4,DTCexp5,DTCexp6,DTCexp7,DTCexp8}, which typically only have access to effective spin-1/2 particles and are limited to tens of particles, can only realize $2T$-periodic DTCs with existing periodic driving schemes. An exception to this is the Nitrogen-Vacancy center experiment of Ref.~\cite{DTCexp2}, which utilizes spin-1 particles and can thus naturally demonstrate $3T$-periodic DTCs. The experimental { observation} of arbitrarily large period DTCs has nevertheless remained a challenge. Accessing these large period DTCs in experiments is however a very important task. Not only will it provide a more definite signature of DTC phases, but it also enables the verification of Mott insulator/Anderson localization in the time domain \cite{DTCcm1,DTCcm2,DTCcm3} and exploration of other unforeseen properties that can only arise in very large period DTCs { \cite{DTCcm8}}. More ambitiously, the ability of realizing large period DTCs on-demand, { especially in the connection with the theory of quantum error correction \cite{QEC1,QEC2,QEC7,QEC8} established here,} may form a first step towards harnessing their potential technological applications, { e.g., as a quantum memory for multiple (possibly highly-entangled) qubits.}



In this paper, we propose an intuitive and realistic approach for constructing an arbitrarily large period DTC with spin-1/2 particles, { the subharmonic signature of which is observable even at small system sizes, e.g., $8-12$ particles, and under a general choice of initial states. As further evidenced by our numerical simulation below, our proposal can thus be immediately confirmed with current successful experimental platforms, such as trapped ions \cite{DTCexp1,DTCexp7} and superconducting circuits \cite{DTCexp8}. In comparison with Refs.~\cite{DTC15,DTC15b}, our proposal utilizes additional realistic interactions beyond the usual Ising-like interaction and operates in the presence disorder, thus representing a construction of MBL-protected (rather than prethermal) large period DTCs.} 

Our proposal is motivated by the striking similarity between DTCs and quantum error correction, both of which exploit a large Hilbert space for protecting features of quantum states encoded in a smaller subspace. Specifically, the Ising-like interaction responsible for protecting observable dynamics in various existing $2T$-periodic DTC models forms a set of stabilizer generators for the well-known quantum repetition codes \cite{Rep1,Rep2,Rep3}. The one-period time-evolution operator (to be referred to as \emph{Floquet operator} onwards) generated by such models can in turn be regarded as a hardware implementation of logical $X$ gate with a quantum repetition code, thus resulting in a robust oscillation between two orthogonal states in a two-dimensional logical subspace. It is natural to expect that, with multiple Ising chains and appropriate interaction among them, one may simulate (within a single driving period) the hardware implementation of various gate operations on several logical qubits arising from multiple quantum repetition codes. In particular, by simulating the specific gate operations proposed below, robust $2^nT$-periodic states can be obtained and detected through the usual average magnetization dynamics. We further show that all the required gate operations can be implemented solely with at most nearest-neighbor two-body interactions, making our proposal particularly attractive for immediate experimental realizations.

\textit{$4T$-periodic DTC.} Before presenting our general construction, we will first elucidate a scheme for generating a $4T$-periodic DTC with two Ising chains to build some intuition. Here, each chain is described by the Hamiltonian $H_{\rm rep,s}=-\sum_j J_{j,s} \mathcal{S}_{j,s,\rm rep}$, where $s=1,2$ and $\mathcal{S}_{j,s,\rm rep}=Z_{j,s}Z_{j+1,s}$ is a stabilizer generator associated with the $s$th chain. Taken together, their collective fourfold degenerate ground state subspace can be spanned by the four logical states $|\overline{0}\overline{0}\rangle \equiv |(0\cdots 0)_1(0\cdots 0)_2\rangle$, $|\overline{0}\overline{1}\rangle \equiv |(0\cdots 0)_1(1\cdots 1)_2\rangle$, $|\overline{1}\overline{0}\rangle \equiv |(1\cdots 1)_1(0\cdots 0)_2\rangle$, and $|\overline{1}\overline{1}\rangle \equiv |(1\cdots 1)_1(1\cdots 1)_2\rangle$. 

It follows that a gate comprising $\overline{X}_1$ followed by $\overline{CNOT}_{1,2}$ ($1$ and $2$ being the control and target qubits respectively) maps $|\overline{0}\overline{0}\rangle\rightarrow |\overline{1}\overline{1}\rangle \rightarrow |\overline{0}\overline{1}\rangle \rightarrow |\overline{1}\overline{0}\rangle \rightarrow |\overline{0}\overline{0}\rangle$. Therefore, evolving any generic state initially prepared in the ground state (or any other fourfold degenerate eigenstate) subspace of $H_{\rm rep,1}+H_{\rm rep,2}$ under the Floquet operator of the form
\begin{equation}
U_T^{(4)} = \overline{CNOT}_{1,2}\overline{X}_1 \times \exp\left(-\mathrm{i} \left[H_{\rm rep,1}+ H_{\rm rep,2}\right]\right) \label{4t1} 
\end{equation} 
gives rise to $4T$ periodicity, detectable by measuring the system's average magnetization. More importantly, if $\overline{CNOT}_{1,2}$ and $\overline{X}_1$ are implemented fault-tolerantly (at least with respect to bit-flip errors, e.g., by the transversal means of Eq.~(\ref{4t2}) below), such $4T$ periodicity is expected to be robust against system imperfections due to a similar error correcting mechanism underlying the quantum repetition codes. In particular, increasing the system size of each Ising chain leads to larger-weight logical $X$ operators, thus further increasing the system resistance against bit-flip errors. Moreover, while repetition codes are generally not capable of correcting phase-flip (Pauli $Z$) errors, which may occur during the imperfect application of transversal $\overline{CNOT}_{1,2}$ gate, we find that such Pauli $Z$ errors do not affect the periodicity of relevant physical states \cite{Supp}.  

\begin{center} 
	\begin{figure}
		\includegraphics[scale=0.45]{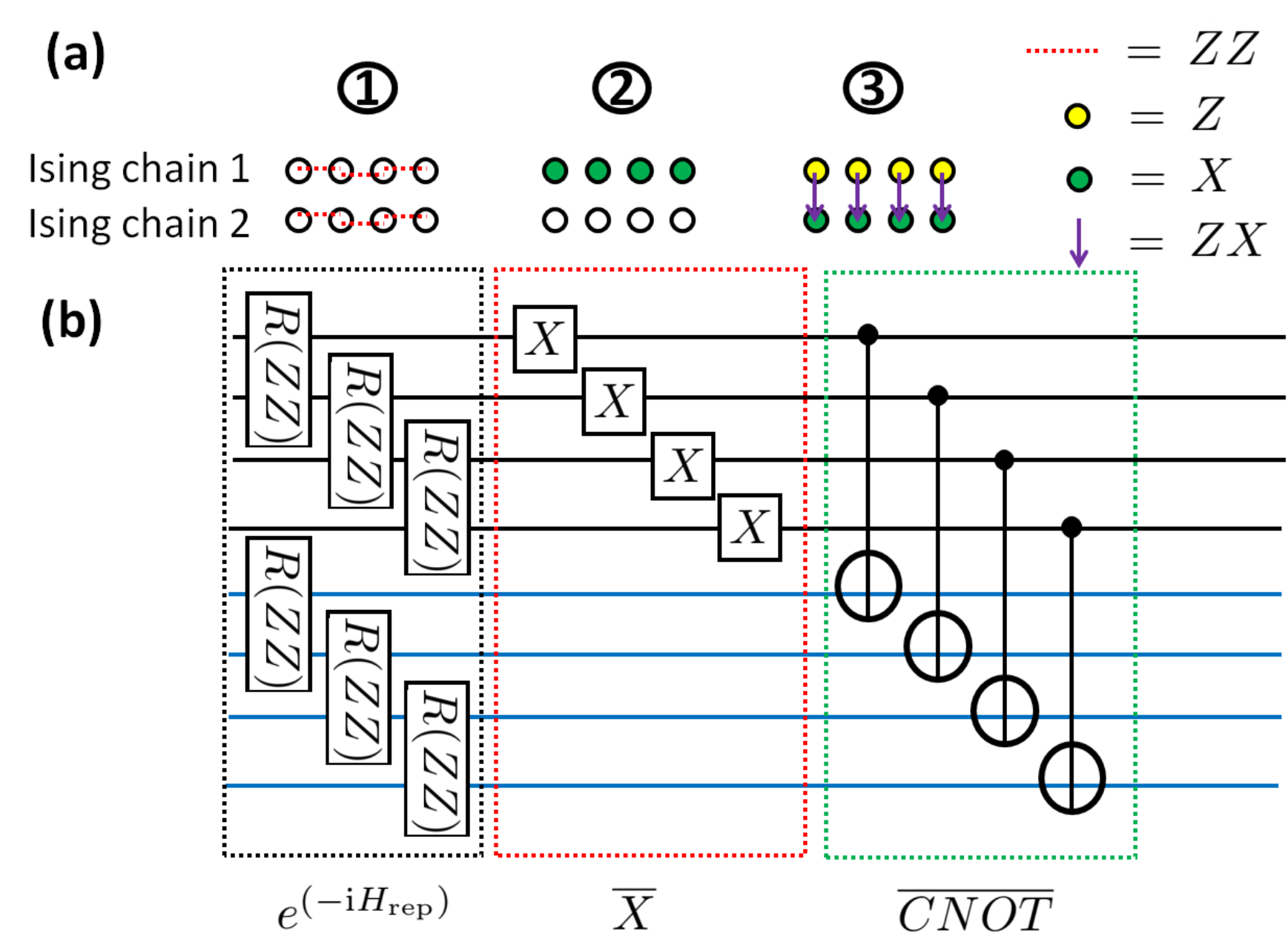}
		\caption{(a) A three-step periodically driven Hamiltonian realizing the Floquet operator Eq.~(\ref{4t1}) with two copies of size-four spin-1/2 Ising chains (each step lasts 1/3rd of the driving period). Spin-spin interactions and single qubit potential are respectively represented by lines and filled circles. (b) The equivalent quantum circuit description of each step, where $R(ZZ)=e^{-\mathrm{i}\theta Z_A Z_B}$ for some (randomly chosen) angle $\theta$.}
		\label{scheme}
	\end{figure}
\end{center}

\begin{center} 
	\begin{figure}
		\includegraphics[scale=0.55]{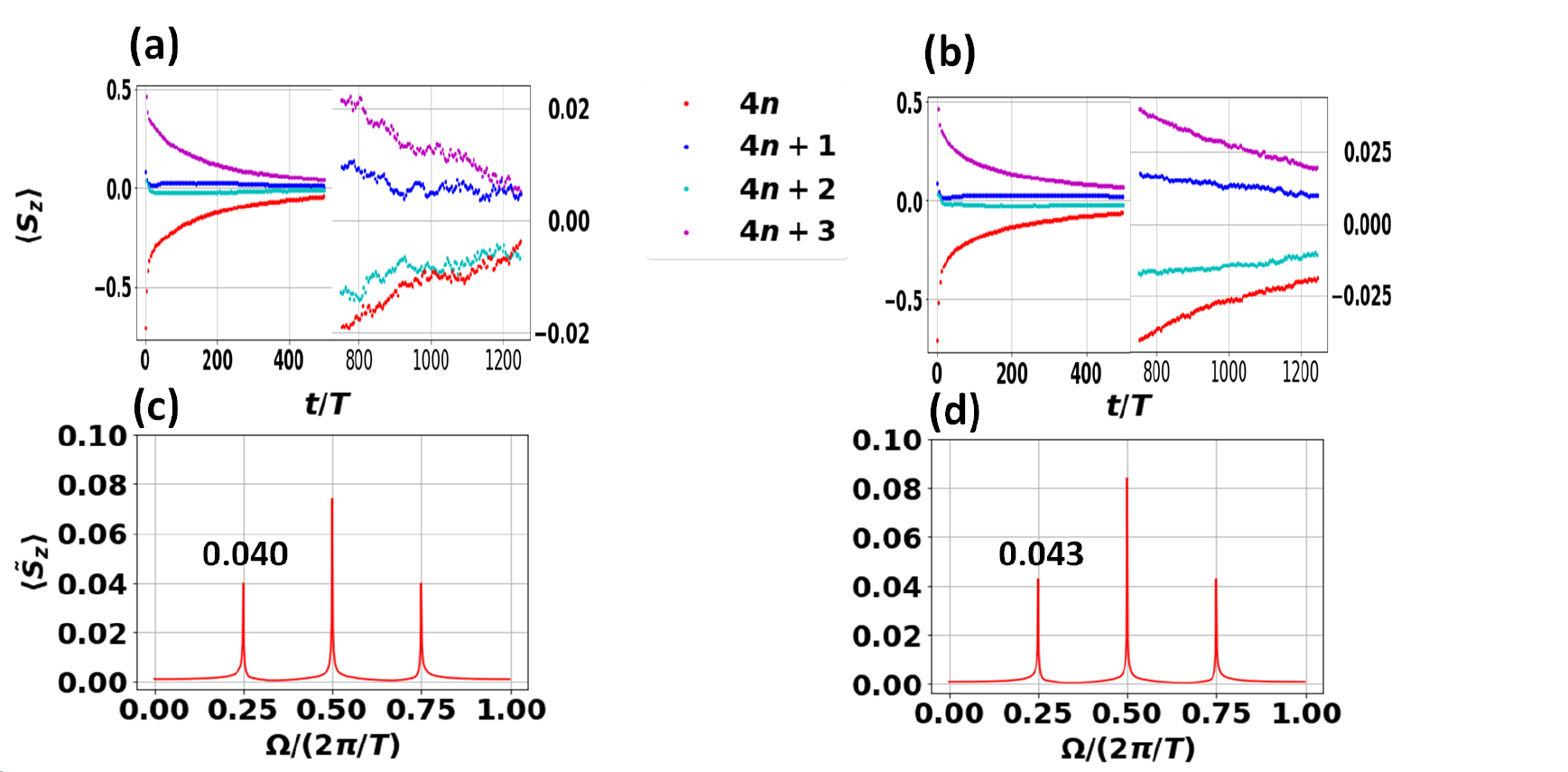}
		\caption{(a,b) Stroboscopic evolution of the average magnetization for a system of two (a) size-four { (8 particles)} and (b) size-five { (10 particles)} spin chains. (c,d) The power spectrum associated with panels (a,b) with respect to the first $500$ Floquet cycles. All system parameters are chosen as $(\bar{J}_1,\Delta J_1)=(1.5,0.5)$, $(\bar{J}_2,\Delta J_2)=(2.5,0.5)$, $(\bar{h}_{j,A},\Delta h_{j,A})=(1.125\pi/2,0.025\pi/2)$, $(\bar{J}_{A,B}^{(ZX)},\Delta J_{A}^{(Z)})=(\bar{J}_{A}^{(Z)},\Delta J_{A}^{(Z)})=(\bar{J}_{B}^{(X)},\Delta J_{B}^{(X)})=(0.925\pi/4,0.025\pi/4)$, and each data point is averaged over $500$ disorder realizations.}
		\label{tworep}
	\end{figure}
\end{center}

To quantitatively demonstrate the above argument, we consider an explicit form of Eq.~(\ref{4t1}) under the transversal implementation
\begin{eqnarray}
\overline{CNOT}_{A,B} &\rightarrow& e^{-\mathrm{i} \sum_{j=1}^N \left(J_{j,A,B}^{(ZX)}Z_{j,A}X_{j,B}+J_{j,A}^{(Z)}Z_{j,A}+J_{j,B}^{(X)}X_{j,B}\right)} \;, \nonumber \\ 
\overline{X}_A &\rightarrow& e^{-\mathrm{i} \sum_{j=1}^N h_{j,A} X_j}\label{4t2} 
\end{eqnarray} 
where $-J_{j,A}^{(Z)}=-J_{j,B}^{(X)}=J_{j,A,B}^{(ZX)}=h_{j,A}/2=\pi/4$ in the ideal case \cite{Supp}. The resulting Floquet operator can, e.g., be obtained from a three-step periodically driven Hamiltonian described schematically in Fig.~\ref{scheme}(a). In Fig.~\ref{tworep}(a,b), we evaluate the stroboscopic time-evolution of the system's average magnetization, i.e., $\langle S_{z} \rangle = \frac{1}{2N} \sum_{j=1}^N \sum_{s=1,2} \langle Z_{j,s} \rangle$. Unless otherwise stated, we take the initial state as $R_{\pi/8}(X)|0\cdots 0 \rangle \equiv\prod_{j=1}^{N}\prod_{s=A,B} e^{-\mathrm{i}\frac{\pi}{8} X_{j,s}}|0\cdots 0\rangle $ in the rest of this paper to emphasize its generality. Moreover, as spatial disorder is necessary for stabilizing DTCs, all parameter values are taken randomly from a uniform distribution of the form $[\bar{P}-\Delta P,\bar{P}-\Delta P]$.

Remarkably, even at considerably large parameter imperfection, i.e., $\gtrapprox 5\%$ from the ideal values implementing the $\overline{X}$ and $\overline{CNOT}_{1,2}$ gates, $4T$-periodic structure is clearly observed and lasts over hundreds Floquet cycles. By evaluating the late-time dynamics of $\langle S_z \rangle$, it follows that the $4T$-periodicity appears to break down after around $1200$ Floquet cycles for the case of size-four chains. By contrast, such $4T$-periodicity remains present for the case of size-five chains, thus confirming the expectation that DTCs possess subharmonic dynamics that persists longer with increase in the system size. Moreover, while the short-time dynamics for both system sizes looks qualitatively similar, its associated power spectrum $\langle \tilde{S}_z \rangle(\Omega) = \left| \frac{1}{\tau}\sum_{j=1}^\tau e^{-\mathrm{i}j \Omega T} \langle S_z \rangle(j) \right|$ reveals sharper subharmonic response at $\Omega = \frac{\pi}{2T},\frac{3\pi}{2T}$ for the case of size-five chains (see Fig.~\ref{tworep}(c,d)). 

\begin{center} 
	\begin{figure}
		\includegraphics[scale=0.4]{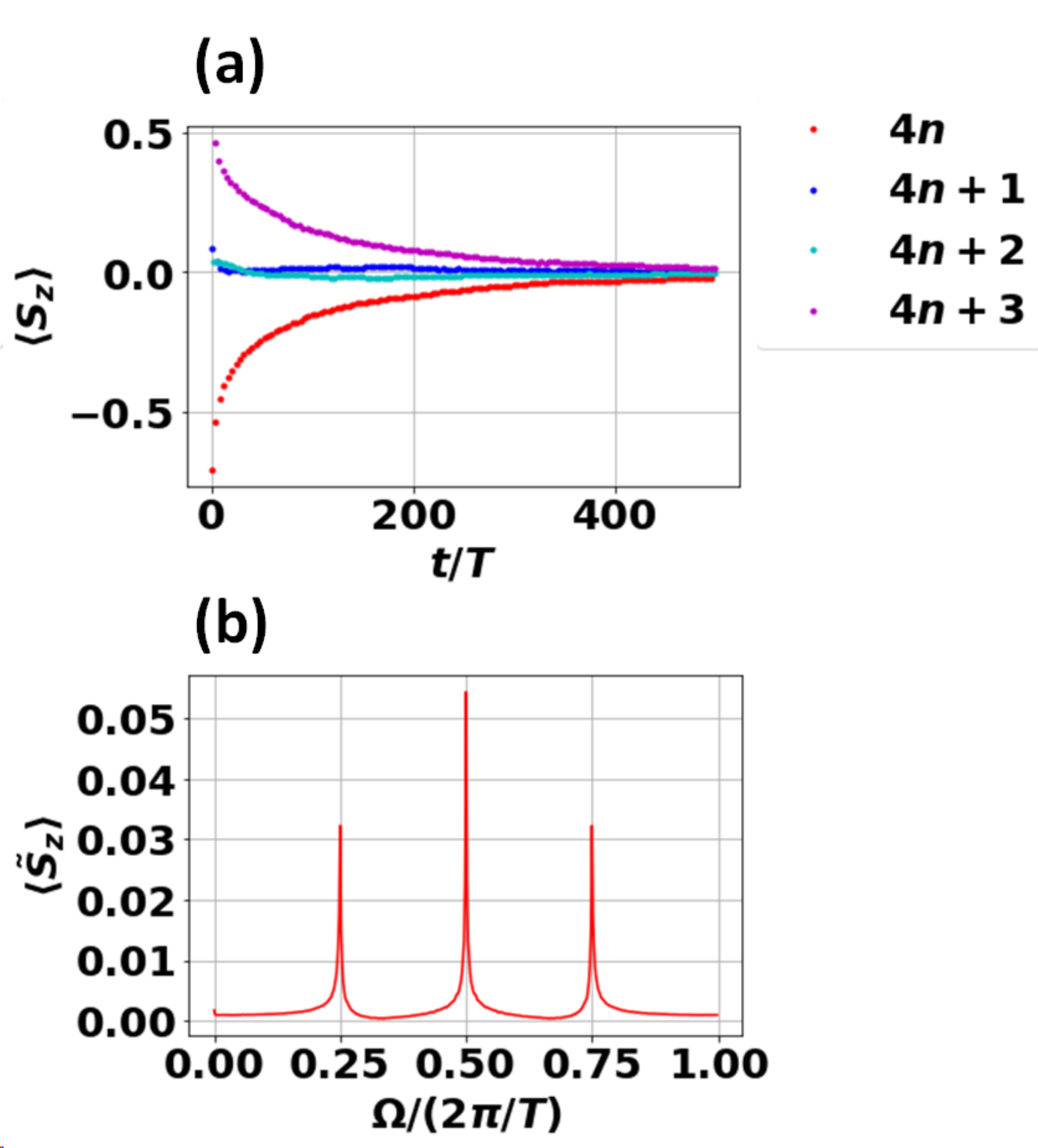}
		\caption{(a) Stroboscopic evolution of the average magnetization for a system of two size-four { (8 particles)} spin chains with long range hopping. (b) The power spectrum associated with panel (a). All system parameters are the same as those of Fig.~\ref{tworep} and each data point is averaged over $500$ disorder realizations.}
		\label{fourTlong}
	\end{figure}
\end{center}

\textit{Experimental simulation of $4T$-periodic DTCs.} A straightforward modification of Eq.~(\ref{4t1}) enables its direct experimental implementation in trapped ions \cite{DTCexp1,DTCexp7} or superconducting qubits (e.g., within the recently emerging Sycamore processor \cite{DTCqs2,DTCexp8,Sycamore}). In trapped ions, effective spin-1/2 particles are obtained from the hyperfine clock states of $^{171}Yb+$ ion, single spin potentials are realized via optically driven Raman transitions between the two states, and (long-range) ZZ interactions can be implemented through spin-dependent optical dipole forces \cite{DTCexp1}. Other spin-spin interactions, e.g., of ZX type, necessary for realizing logical $CNOT$ gates can also be obtained by applying appropriate $\pi/2$ pulses to the more natural ZZ interactions. Note that such $\pi/2$ pulses have also been explicitly used in Ref.~\cite{DTCexp1} to realize a disordered $Z$ field. To verify the observability of our $4T$-periodic DTC in trapped ions \cite{DTCexp1,DTCexp7}, we modify 
\begin{equation}
H_{\rm rep,s} \rightarrow \sum_{j=2}^N \sum_{k<j} J_{j,k,s} \frac{Z_{j,s} Z_{k,s}}{|j-k|^{1.5}}
\end{equation}
in Eq.~(\ref{4t1}), which establishes its compatibility with the long-range nature of the spin-dependent optical forces. The simulated average magnetization dynamics in Fig.~\ref{fourTlong} reveals that the expected $4T$-periodicity indeed remains present in such a system.

\begin{center} 
	\begin{figure}
		\includegraphics[scale=0.4]{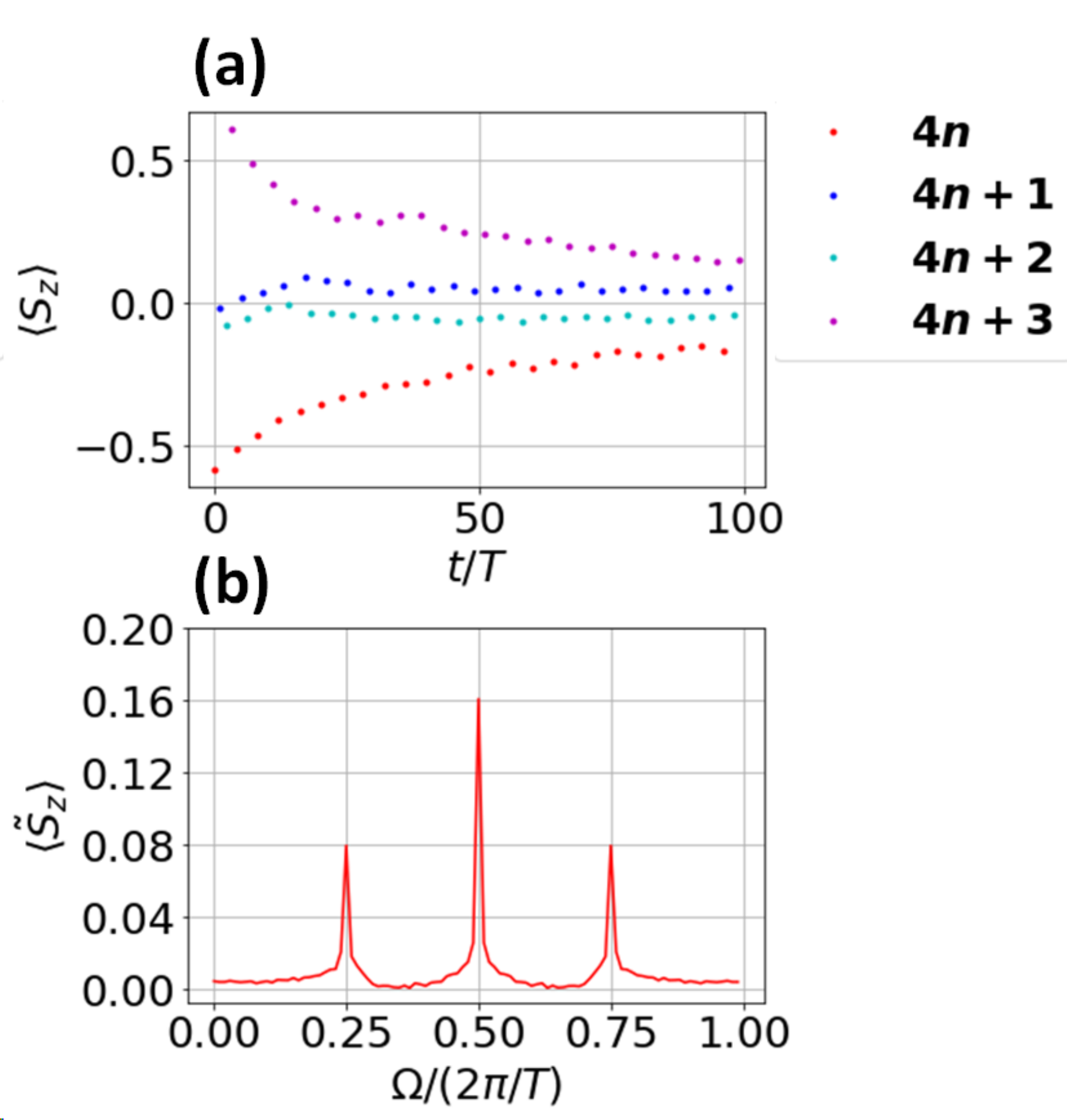}
		\caption{(a) Sycamore's simulated stroboscopic evolution of a single qubit magnetization under repeated application of the circuit simulating Eq.~(\ref{4t1}) (See \cite{Supp} for detail). The state is initialized in $\prod_{j=1}^N \prod_{s=1,2} e^{-\mathrm{i} \theta_{\pi/8} X_{j,s}} |0\cdots 0\rangle${, where $N=16$,} by applying imperfect $\theta_{\pi/8}\approx \pi/8$ Pauli $X$ rotation on all qubits at the beginning of the simulation. (b) The power spectrum associated with panel (a). In all panels, we take $(\bar{J}_1,\Delta J_1)=(1.5,0.5)$, $(\bar{J}_2,\Delta J_2)=(2.5,0.5)$}
		\label{cirsyc2}
	\end{figure}
\end{center}

In the superconducting qubits, Eq.~(\ref{4t1}) can be simulated directly at the unitary level. In the following, we shall focus on the Google's recent successful Sycamore experiment \cite{DTCexp8,Sycamore}, but we expect that our scheme can also be realized in other superconducting settings. In this case, the time-evolution associated with on-site potentials is simulated with appropriate single-qubit rotation, whereas that associated with nearest-neighbor ZZ or ZX interaction is created by conjugating a single-qubit rotation with the iSWAP gate. Further detail, along with the complete circuit simulating Eq.~(\ref{4t1}), is presented in \cite{Supp}. Figure~\ref{cirsyc2} shows the simulated stroboscopic single qubit magnetization obtained via utilizing Python's CIRQ package \cite{CIRQ}. Specifically, after inputting an initial state on a quantum circuit with $16$ qubits and simulating $n$ repeated applications of $U_T^{(4)}$, we obtain the single qubit magnetization at time $nT$ by measuring the qubit in the $Z$ basis and repeating the experiment $480$ times. This procedure is further repeated over $20$ disorder realizations of the system parameters. To better simulate realistic conditions, we have assumed that, in addition to intentional imperfection in the system parameters enacting the $\overline{X}$ and $\overline{CNOT}$ gates by up to $7.5\%$, each implementation of single-qubit rotation and iSWAP gate is imperfect with error of up to $0.5\%$ and $4\%$ respectively. It is remarkable to note that $4T$-periodicity persists even in the presence of such temporal errors, thus highlighting the feasibility of observing our proposed DTC features in the actual Sycamore experiment.

\begin{center} 
	\begin{figure}
		\includegraphics[scale=0.45]{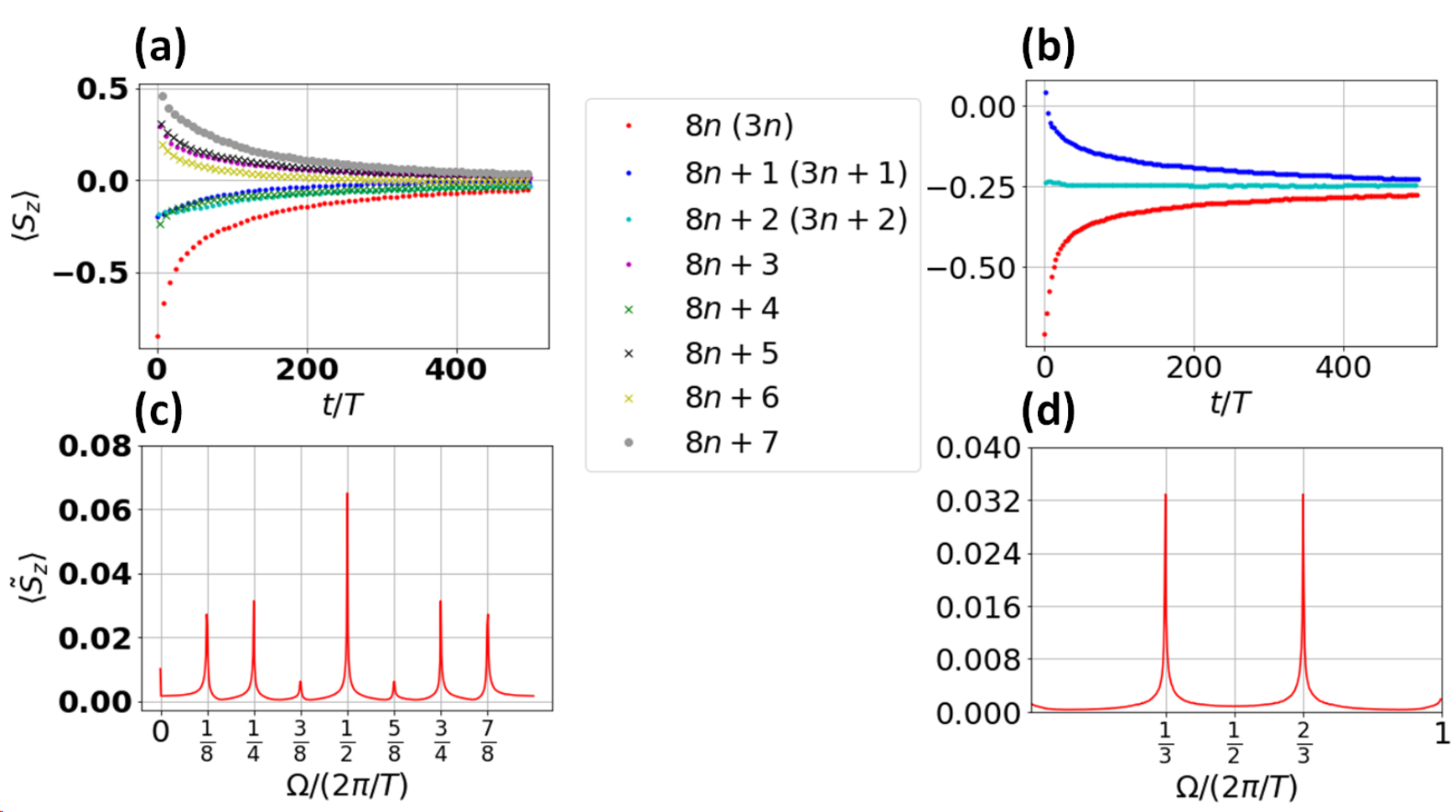}
		\caption{(a,b) Stroboscopic evolution of the averaged magnetization { in three arrays of size-four (12 particles) spin chains} under imperfect (a) $U_T^{(8)}$, (b) $U_T^{(3)}$. (c, d) The power spectrum associated with panels (a, b). In all panels, we take $(\bar{J}_1,\Delta J_1)=(1,0.5)$, $(\bar{J}_2,\Delta J_2)=(1.5,0.5)$, $(\bar{J}_2,\Delta J_2)=(2,0.5)$ and average each data point over $100$ disorder realizations. Other system parameters for simulating $\overline{X}$, $\overline{CNOT}$, and $\overline{CCNOT}$ deviate from their ideal values with an error randomly chosen between $5-10\%$.}
		\label{threerep}
	\end{figure}
\end{center}

\textit{$8T$- and $3T$-periodic DTCs.} To demonstrate another nontrivial example of large-period DTCs, we now consider three arrays of size-$N$ spin chains, which are subjected to a time-periodic Hamiltonian associated with the Floquet operator
\begin{eqnarray}
U_T^{(8)} &=& \overline{CCNOT}_{12,3} \overline{CNOT}_{1,2} \overline{X}_1 \times e^{-\mathrm{i} \sum_{s=1}^3 H_{{\rm rep},s}}\;, \nonumber \\ \label{rep3}
\end{eqnarray}
where $\overline{CNOT}_{A,B}$ and $\overline{X}_A$ are implemented according to Eq.~(\ref{4t2}), while $\overline{CCNOT}_{AB,C}$ can also be implemented transversally as \cite{Supp}
\begin{eqnarray}
\overline{CCNOT}_{AB,C} &=& e^{-\sum_{j=1}^N\frac{\mathrm{i}\pi}{8}(1-Z_{j,1})(1-Z_{j,2})(1-X_{j,3})} \;. \label{8T3}
\end{eqnarray}
In this case, $U_T^{(8)}$ maps the eight logical basis states $|\overline{s}_1\overline{s}_2 \overline{s}_3\rangle$, where $\overline{s}_1,\overline{s}_2,\overline{s}_3=\overline{0},\overline{1}$, into one another to yield an $8T$-periodicity. In Fig.~\ref{threerep}(a), we plot the stroboscopic evolution of $\langle S_z \rangle$ under a considerably imperfect implementation of $U_T^{(8)}$, i.e., $\gtrapprox 5\%$ deviation on all system parameters from their ideal values associated with perfect $\overline{CNOT}$ and $\overline{CCNOT}$ gates.  

Remarkably, devising a time-periodic scheme that simulates another set of logical gates may yield DTCs with different subharmonic dynamics. For instance, consider the Floquet operator
\begin{equation}
U_T^{(3)} = \overline{CNOT}_{2,1} \overline{CNOT}_{1,2} \overline{CNOT}_{3,2} \overline{X}_1 \times e^{\left(-\mathrm{i} \sum_{s=1}^3 H_{{\rm rep},s}\right)} \;. \label{(rep3T)}
\end{equation} 
It is easily checked that $U_T^{(3)}$ maps $|\overline{000}\rangle \rightarrow |\overline{010}\rangle \rightarrow |\overline{100} \rangle \rightarrow |\overline{000} \rangle$ and $|\overline{111}\rangle \rightarrow |\overline{001}\rangle \rightarrow |\overline{101} \rangle \rightarrow |\overline{111} \rangle$, both generating a $3T$-periodicity. In Fig.~\ref{threerep}(b), we verify the robustness of the expected $3T$-periodic signature in $\langle S_z\rangle$ under the imperfect application of $U_T^{(3)}$.

\textit{$2^nT$-periodic DTC.} Generalizing the idea from the previous examples, we will now present a scheme to systematically construct any $2^nT$-periodic DTCs. To this end, one may utilize $n$ copies of spin-1/2 chains and apply a time-periodic Hamiltonian that leads to the Floquet operator
\begin{equation}
U_T^{(2^n)} = \left(\prod_{j=1}^{n-1} \overline{C^{(j)}NOT}_{1\cdots j, j+1} \right) \overline{X}_1 e^{\left(-\mathrm{i}\sum_{s=1}^n H_{{\rm rep},s}\right)} \;, \label{genDTC}
\end{equation}
where $C^{(j)}NOT_{1\cdots j, j+1}=e^{\left[-\mathrm{i} \frac{\pi}{2^{j+1}}\left(\prod_{k=1}^j (1-Z_k)\right) (1-X_{j+1})\right]}$ is the generalized $CNOT$ gate. Intuitively, the gates enacted by $U_T^{(2^n)}$ realize a subtractive operation modulo $2^n$ on any logical state in the computational basis. That is, by writing $|\overline{j}\rangle \equiv |\overline{s}_1\cdots \overline{s}_n \rangle$ such that $\overline{j}=\left(\sum_{k=0}^{n-1} \overline{s}_{k+1} 2^k \right) \; \mathrm{mod} \; 2^n$, it follows that $\left(U_T^{(2^n)}\right)^k |\overline{j} \rangle = |\overline{j-k} \rangle$ for $k<2^n$ and $\left(U_T^{(2^n)}\right)^{2^n} |\overline{j} \rangle = |\overline{j} \rangle$, thus establishing the desired $2^nT$ periodicity. In \cite{Supp}, we use Floquet theory \cite{Flo1,Flo2} to further elucidate the robustness of such $2^nT$ periodicity against parameter imperfection. It is worth noting that such robustness, as well as the formulation in the language of quantum repetition code, suggests the potential of large-period DTCs as a quantum memory for highly-entangled qubits. Indeed, one can in principle store such qubits in the logical subspace spanned by the states $|\overline{j}\rangle$.

Apart from the $n=2$ case considered above, $U_T^{(2^n)}$ may not appear to be realistic due to the presence of various nonlocal and higher-weight spin-spin interactions in the transversal implementation of $\overline{C^{(j)}NOT}_{1\cdots j, j+1}$. As we show in \cite{Supp}, however, one can in principle replace all these non-realistic interactions by a series of nearest-neighbor weight-two spin-spin interactions, at the expense of causing a single Floquet cycle to be potentially longer. Alternatively, in the spirit of $U_T^{(3)}$ of Eq.~(\ref{(rep3T)}), it might be possible to realize $2^nT$- or potentially any $nT$-periodic DTCs with simpler local Floquet operator by utilizing $n'>n$ copies of spin chains. We leave the exploration of this opportunity as a potential future work. These systems can in turn be readily adapted for experimental implementation in either trapped ions (by replacing every nearest-neighbor interaction with a Coulomb-like interaction) or superconducting circuits (by simulating the time-evolution of every nearest-neighbor interaction with a series of single-qubit and iSWAP gates).  

\textit{Concluding Remarks.} We have proposed a general scheme for building large period DTCs that are observable even at small number of spin-1/2 particles. In particular, having demonstrated the explicit construction and strong signatures of $4T$- and $8T$-periodic DTCs, we elucidate how an arbitrarily large $2^nT$-periodic DTC can be constructed with at least $n$-copies of spin-1/2 chains, under a periodic driving scheme that simulates appropriate logical gate operations on quantum repetition codes. Most importantly, we highlight the possibility of implementing the required periodic driving scheme using only at most nearest-neighbor weight-two spin-spin interactions, thus allowing for its experimental realization in the immediate future. Indeed, explicit numerical simulation of trapped ions and Sycamore experiments has also been carried out for our $4T$-periodic DTC.

Moving forward, with the availability of other more sophisticated quantum error correction codes, e.g., the surface \cite{QEC3,QEC4,QEC5,QEC6,QEC9,QEC10} and color codes \cite{QEC11,QEC12}, applying the same strategy above opens promising opportunies for discovering new Floquet phases of matter \cite{DTCSC1,DTCSC2,DTCSC3}. To this end, one simply needs to devise two types of Hamiltonians over one driving period, i.e., one that simulates the codes' stabilizer generators, as well as one that simulates the transversal implementation of some logical gate operations. Even within the area of DTCs, the realization of large-period DTCs should enable the experimental exploration of exotic properties not accessible within $2T$-periodic setting, such as the analogues of Anderson localization and Mott insulator in the time domain \cite{DTCcm2,DTC25}, as well as time-induced many-body localization \cite{DTCcm5}.

\begin{acknowledgments}
	{\bf Acknowledgement}: This work is supported by the Australian Research Council Centre of Excellence for Engineered Quantum Systems (EQUS, CE170100009). The author thanks Jiangbin Gong and Stephen Bartlett for reading the first draft of this manuscript and providing valuable comments. The author also thanks Benjamin Brown for providing useful feedback.
\end{acknowledgments}

\newpage
\onecolumngrid 
\appendix

\begin{center}
	\textbf{Quantum repetition codes as building blocks of large period discrete time crystals: Supplemental Material}
\end{center}

This Supplemental Material consists of five sections. In Sec.~A, we discuss the origin and robustness of $2^nT$ magnetization dynamics in the proposed $2^nT$-periodic DTCs from the perspective of Floquet theory and quantum error correction. In Sec.~B, we review how quantum gate operations $X_j$ and $C^{(k)}NOT_{1\cdots k,k+1}$ operate. In Sec.~C, we show that the $\overline{C^{(k)}NOT}_{1\cdots k,k+1}$ gate can be implemented transversally. We further show that such a gate can also be implemented with at most weight-two nearest-neighbor interactions in Sec.~D. Finally, Sec.~E elucidates in detail the Sycamore circuit implementation of the $4T$-periodic DTC model presented in the main text.

\section{Section A: Robustness of $2^{n}T$-periodic DTCs}
\label{app0}

\subsection{Overview of 2T-periodic DTCs}

To develop an intuition into the connection between DTCs and the quantum repetition code, we will first overview a $2T$ periodic DTC model described by the Floquet operator \cite{DTC2}, 
\begin{equation}
U_T = \exp\left(-\mathrm{i} \sum_{j=1}^N h_j X_j \right) \exp\left(\mathrm{i} \sum_{j=1}^N \left[J_j Z_j Z_{j+1}+h_j^{(Z)} Z_j\right] \right) \;, \label{repetition}
\end{equation}
where $j$ and $N$ represent the lattice site and system size respectively, $P_j$ with $P\in\mathrm{X,Y,Z}$ are the Pauli operators at site $j$, $J_j$ is a site-dependent Ising interaction, and $h_j$ is proportional to the external magnetic field that is tuned near, but may not be exactly at, $\pi/2$. Here, the terms $h_j Z_j$ serve both as a disordered part that contributes towards establishing many-body localization (MBL), as well as to make Eq.~(\ref{repetition}) effectively interacting. 

From the perspective of quantum error correction, the terms $J_j Z_j Z_{j+1}$ in the second exponential of Eq.~(\ref{repetition}) can be identified as a Hamiltonian associated with a quantum repetition code, i.e., $H_{\rm rep} = -\sum_{j=1}^N J_j \mathcal{S}_{j,\rm rep}$, where $\mathcal{S}_{j,\rm rep}=Z_jZ_{j+1}$ is a set of stabilizer generators (mutually commuting operators) describing such a code. In this case, the associated logical operators are identified as $\overline{Z}=Z_1$ and $\overline{X}=\prod_{j=1}^N X_j$. Two degenerate ground states of $H_{\rm rep}$ that are also simultaneous eigenstates of $\overline{Z}=\pm 1$ and $\mathcal{S}_{j,\rm rep}=+1$ can be obtained as $|\overline{0}\rangle=|0\cdots 0\rangle$ and $|\overline{1}\rangle=|1\cdots 1\rangle$. Moreover, at special parameter values $h_j=\pi/2$, the first exponential in Eq.~(\ref{repetition}) reduces exactly to $\overline{X}$. Consequently, a state initially prepared in $|\overline{0}\rangle$ will transform to $|\overline{1}\rangle$ after one period and vice versa. In the language of Floquet theory \cite{Flo1,Flo2}, $\overline{X}$ splits the degenerate logical subspace of the $H_{\rm rep}$ into two nondegenerate Floquet eigenstate subspace of $U_T(h_j=\pi/2) \equiv U_{T,\rm ideal}$ with quasienergy (phase of eigenvalue) separation of exactly $\pi/T$. Explicitly, two such Floquet eigenstates can be exactly written as 
\begin{equation}
|\varepsilon_{0,\pm}\rangle = e^{-\sum_{j=1}^N h_j^{(Z)}} |\overline{0}\rangle \pm e^{\sum_{j=1}^N h_j^{(Z)}} |\overline{1}\rangle \;. \label{2tflo}
\end{equation}
Consequently, $|\overline{0}\rangle$ and $|\overline{1}\rangle$ now represent a linear combination of $|\varepsilon_{0,\pm}\rangle$ which pick up a relative phase of $\pi$ after one period and are $2T$ periodic. Such $2T$ periodicity then manifests itself as the subharmonic dynamics of some observable, e.g., the average magnetization. 

As $h_j$ is tuned away from $\pi/2$, Eq.~(\ref{repetition}) can be understood as evolving the system under the ideal logical $\overline{X}$, followed by a series of small Pauli $X$ rotations acting on all the physical qubits. Mathematically, one may write,
\begin{equation}
U_T = U' U_{T,\rm ideal} = e^{-\mathrm{i} \sum_{j=1}^N \epsilon_j X_j} \overline{X} e^{\mathrm{i} \sum_{j=1}^N \left( J_j Z_j Z_{j+1} + h_j^{(Z)} Z_j \right)} \;.
\end{equation}
A state initially prepared in $|\overline{0}\rangle$ will thus evolve to 
\begin{eqnarray}
U' |\overline{1}\rangle &=& \langle \varepsilon_{0,+} | U' |\overline{1}\rangle |\varepsilon_{0,+}\rangle+ \langle \varepsilon_{0,-} | U' |\overline{1}\rangle |\varepsilon_{0,-}\rangle + \cdots \;,
\end{eqnarray} 
where $(\cdots)$ contains a summation over all other Floquet eigenstates of $U_{T,\rm ideal}$. By expanding $U'$ and assuming that all $\epsilon_j$ are of the same order, i.e., $\sin(\epsilon_j)\sim \epsilon \ll 1$, it is easily checked that $\mathrm{i} \; \mathrm{log}\left(\frac{\langle \varepsilon_{0,+} | U' |\overline{1}\rangle }{\langle \varepsilon_{0,-} | U' |\overline{1}\rangle }\right)=\pi+\mathcal{O}\left(\epsilon^N\right)$. That is, up to a very small correction of order $\epsilon^N$, which arises from the $X_1\cdots X_n$ term of $U'$, the resulting state after one period can approximately be written as a sum of two Floquet eigenstates with a relative phase of $\pi$. In this case, a physical error of order $\epsilon$ in the implementation of logical $\overline{X}$ is suppressed to an error of order $\epsilon^N$ in the actual observable dynamics, thus highlighting the error correction mechanism of $2T$ periodic DTCs. 

Before ending this section, we note that the above argument can be extended to other stabilizer subspaces of $H_{\rm rep}$ to arrive at the conclusion that any product state of physical $|0\rangle$ and $|1\rangle$ exhibits robust $2T$ periodicity. It is however also important to note that we have implicitly assumed the existence of finite energy gap between different stabilizer subspaces of $H_{\rm rep}$. In the presence of accidental degeneracy between some stabilizer subspaces, certain perturbations may yield a Floquet operator whose eigenstates contain superposition of states belonging to different stabilizer subspaces. In this case, the system may lose its ability to suppress error to $N$-th order since lower weight Pauli $X$ operators may now yield finite overlap between two relevant eigenstates. Such detrimental accidental degeneracy can be removed by ensuring strong disorder in the repetition code Hamiltonian. This in turn implies the necessity of disorder in all the systems proposed in the main text.

\subsection{Generalization to $2^nT$-periodic DTCs}

As intuitively elucidated in the main text, $U_T^{(2^n)}$ is designed so that it cyclically transforms the $2^n$ computational basis states in the logical subspace into one another. To understand how error correction mechanism protects such $2^nT$ periodicity against parameter imperfection and variation in the initial state, it is important to first understand its origin from the perspective of Floquet theory. Generalizing the argument presented in the previous section, we will first assume that all $\overline{C^{(j)}NOT}_{1,\cdots j, j+1}$ and $\overline{X}_1$ appearing in Eq.~(7) of the main text are implemented ideally. In this case, the system becomes integrable, where all of its Floquet eigenstates and their associated quasienergies can be analytically obtained.

We start by again noting that $\sum_{s=1}^n H_{\rm rep,s}$ consists of a sum of mutually commuting stabilizer operators $-J_{j,s}\mathcal{S}_{j,s}=-J_{j,s}Z_{j,s}Z_{j+1,s}$. Its eigenstates can thus be easily found by diagonalizing $\mathcal{S}_{j,s}=S_{j,s}=\pm 1$ and are at least $2^n$-fold degenerate with energy $E_{\left\lbrace S_{j,s}\right\rbrace}=-\sum_{j=1}^N \sum_{s=1}^n J_{j,s} S_{j,s}$. In particular, the computational basis states $|\overline{s}_1 \cdots \overline{s}_n \rangle$, where each $\overline{s}_1,\overline{s}_2,\cdots,\overline{s}_n=0,1$ defines the logical state of each repetition code (Ising spin) Hamiltonian, form a $2^n$-fold degenerate ground state subspace at energy $E_{0}=-\sum_{j=1}^N \sum_{s=1}^n J_{j,s}$. While the ideal logical gate operation $\left(\prod_{j=1}^{n-1} \overline{C^{(j)}NOT}_{1\cdots j, j+1}^{(\rm ideal)} \right) \overline{X}_1^{(\rm ideal)}$ lifts such a degeneracy, it can be easily checked that the linear combinations
\begin{equation}
|\varepsilon_{0,\ell}\rangle = \sum_{j=0}^{2^{n}-1}  e^{\mathrm{i} \frac{j \pi \ell}{2^{n-1}} } |\overline{j}\rangle \;,
\end{equation}
where $\ell=0,\cdots,2^{n}-1$ and $|\overline{j}\rangle \equiv |\overline{s}_1 \cdots \overline{s}_n \rangle$ such that $\overline{j}=\left(\sum_{k=0}^{n-1} \overline{s}_{k+1} 2^k \right) \; \mathrm{mod} \; 2^n$, remains a Floquet eigenstate of $U_{T,\rm ideal}^{(2n)}$ with quasienergy
\begin{equation}
\varepsilon_{0,\ell} = \left(E_0 -\frac{\ell \pi}{2^{n-1} T} \right) \; \mathrm{mod} \; \frac{2\pi}{T} \;.
\end{equation}  
That is, $|\varepsilon_{0,\ell}\rangle $ form a set of Floquet eigenstates with $\frac{\pi}{2^{n-1}T}$ quasienergy separation. Moreover, since
\begin{equation}
|\overline{j} \rangle = \sum_{\ell=0}^{2^{n}-1}  e^{-\mathrm{i} \frac{j \pi \ell}{2^{n-1}} } |\varepsilon_{0,\ell} \rangle \;, \label{2n}
\end{equation}
it follows that $(U_{T,\rm ideal}^{(2^n)})^{2^n} |\overline{j} \rangle \propto |\overline{j} \rangle$ but $(U_{T,\rm ideal}^{(2^n)})^{k} |\overline{j} \rangle \propto |\overline{(j-k)} \rangle \neq |\overline{j} \rangle$ for $k<2^n$ due to the relative phase of $\frac{\pi}{2^{n-1}}$ among all $|\varepsilon_{0,\ell} \rangle$ at each period. The above argument can be straightforwardly repeated with respect to other degenerate eigenstate subspaces of $\sum_{s=1}^n H_{\rm rep,s}$ to show that \emph{any} product state of physical $|0\rangle$ and $|1\rangle$ is $2^n T$-periodic. Remarkably, as evidenced by the various numerical results presented in the main text, such $2^n T$-periodicity in fact also holds for other product states of any arbitrary $|\psi\rangle\equiv a|0\rangle +b|1\rangle $ for some normalized constants $a$ and $b$.     

We now note that, since $C^{(j)}NOT_{1\cdots j, j+1}=e^{\left[-\mathrm{i} \frac{\pi}{2^{j+1}}\left(\prod_{k=1}^j (1-Z_k)\right) (1-X_{j+1})\right]}$ and each term in the exponential commutes with one another, we may write any imperfect application of logical $C^{(j)}NOT_{1\cdots j,j+1}$ as
\begin{equation}
\overline{C^{(j)}NOT_{1\cdots j,j+1}} =\widetilde{C^{(j)}NOT_{1\cdots j,j+1}} \overline{C^{(j)}NOT_{1\cdots j,j+1}}^{\rm (ideal)} \;,  
\end{equation} 
where $\overline{C^{(j)}NOT_{1\cdots j,j+1}}^{\rm (ideal)}$ is the ideal application of logical $C^{(j)}NOT_{1\cdots j,j+1}$ (see Sec.~B for a review of its definition), whereas $\widetilde{C^{(j)}NOT_{1\cdots j,j+1}}$ reflects its imperfection. In this case, Eq.~(7) in the main text can be written as
\begin{eqnarray}
U_T^{(2^n)}	&=& \widetilde{C^{(n-1)}NOT}_{1\cdots n-2,n-1} \widetilde{C^{(n-2)}NOT}_{1\cdots n-3,n-2}' \times \cdots \times  \widetilde{CNOT}_{1,2}^{'(n-1)} \widetilde{X}_1^{'(n)} \nonumber \\
& \times & \left(\prod_{j=1}^{n-1} \overline{C^{(j)}NOT}_{1\cdots j, j+1}^{(\rm ideal)} \right) \overline{X}_1^{(\rm ideal)} e^{\left(-\mathrm{i}\sum_{s=1}^n H_{{\rm rep},s}\right)} \nonumber \\
&=& \widetilde{U}_{T}^{(2^n)} U_{T,\rm ideal}^{(2^n)} \;,       
\end{eqnarray}
where we have defined
\begin{equation}
\widetilde{P}^{'(j)} = \left(\prod_{k=n-j}^{n-1} \overline{C^{(k)}NOT}_{1\cdots k-1, k}^{(\rm ideal)}  \right) \widetilde{P} \left(\prod_{k=n-j}^{n-1} \overline{C^{(k)}NOT}_{1\cdots k-1, k}^{(\rm ideal)}  \right)^{(-1)} \;.
\end{equation}

Consider again the product state $|\overline{j}\rangle$ defined in Eq.~(\ref{2n}). After a single period, it evolves to 
\begin{eqnarray}
\widetilde{U}_{T}^{(2^n)} U_{T,\rm ideal}^{(2^n)} |\overline{j}\rangle &=& \sum_{\ell' = 0}^{2^n-1} e^{-\mathrm{i} E_0 T} \langle \varepsilon_{0,\ell'} | \widetilde{U}_{T}^{(2^n)} |\overline{(j-1)} \rangle |\varepsilon_{0,\ell'} \rangle \nonumber \\
&+& \sum_{m\neq 0} \sum_{\ell'=0}^{2^n-1} e^{-\mathrm{i} E_0 T} \langle \varepsilon_{m,\ell'} | \widetilde{U}_{T}^{(2^n)} |\overline{(j-1)} \rangle |\varepsilon_{m,\ell'} \rangle \;, \label{check}
\end{eqnarray}
In general, $\widetilde{U}_{T}^{(2^n)} |\varepsilon_{0,\ell} \rangle$ now has support on all eigenstates of $U_{T,\rm ideal}^{(2^n)}$, i.e., $\langle \varepsilon_{m,\ell'} | \widetilde{U}_{T}^{(2^n)} |\varepsilon_{0,\ell} \rangle \neq 0$. For the purpose of highlighting the robustness of the $2^nT$ periodicity, however, it is sufficient to show that its support on eigenstates $|\varepsilon_{0,\ell}\rangle$ maintains a relative phase of $\approx \frac{\pi}{2^{n-1}}$. By construction, $\tilde{U}_T^{(2^n)}$ consists of products of Pauli exponentials, which can be expanded into a superposition of products of Pauli operators. Note in particular that low-weight Pauli $X$ operators bring $|\overline{(j-1)}\rangle$ to a state outside the logical subspace, which has zero overlap with any $|\varepsilon_{0,\ell}\rangle$. For Ising chain of size $N$, Pauli $X$ operators with at least weight-$N$ are necessary for yielding a finite overlap between $|\overline{(j-1)}\rangle$ with some $|\varepsilon_{0,\ell'}\rangle$. Provided that all system imperfections are of order $\epsilon$, such large weight Pauli $X$ operators only appear in $\widetilde{U}_T^{(2^n)}$ as terms of order at least $\epsilon^N$. Terms involving Pauli $Z$ operators may at first appear more dangerous as they preserve the logical subspace containing $|\varepsilon_{0,\ell}\rangle$. However, note that $\langle \varepsilon_{m,\ell'} | Z_{k,s} |\overline{(j-1)} \rangle = e^{-\mathrm{i} \frac{(j-1)\pi \ell'}{2^{n-1}T}} $ if $0\leq j-1< 2^s$, while $\langle \varepsilon_{m,\ell'} | Z_{k,s} |\overline{(j-1)} \rangle = -e^{-\mathrm{i} \frac{(j-1)\pi \ell'}{2^{n-1}T}} $ if $ 2^s\leq j-1<2^n $. In both cases, Pauli $Z$ contributions to $\widetilde{U}^{(2^n)}$ preserve the relative phase of $\approx \frac{\pi}{2^{n-1}}$ among all $|\varepsilon_{0,\ell}\rangle$ and will thus not destroy the $2^n T$ periodicity.   

\section{Section B: Review of quantum gate operations $X_j$ and $C^{(k)}NOT_{1\cdots k,k+1}$}       

Given $n\geq j, k+1$ qubits spanned by the states $|s_1 s_2 \cdots s_n \rangle $ in the computational basis with $s_1,s_2,\cdots, s_n=0,1$, the Pauli $X_j$ operates as 
\begin{equation}
X_j |s_1 s_2\cdots s_n\rangle = |s_1 \cdots s_{j-1} (1-s_j) s_{j+1} \cdots s_n\rangle \;,
\end{equation}
whereas $C^{(k)}NOT_{1\cdots k,k+1}$ gate operates as 
\begin{equation}
C^{(k)}NOT_{1\cdots k,k+1} |s_1 s_2\cdots s_n\rangle = \begin{cases}
|s_1 s_2\cdots s_k (1-s_{k+1}) s_{k+2} \cdots s_n\rangle & \text{ if } s_1,\cdots, s_k =1 \\
|s_1 s_2\cdots s_k s_{k+1} s_{k+2} \cdots s_n\rangle & \text{ otherwise }
\end{cases} \;.
\end{equation}
The logical quantum gates $\overline{X}_j$ and $\overline{C^{(k)}NOT}_{1\cdots,k,k+1}$ operate similarly on the logical quantum state $|\overline{s}_1 \cdots \overline{s}_n\rangle $, where each $|\overline{s}_\ell \rangle$ now consists of multiple physical qubits.

\section{Section C: Transversal implementation of $\overline{C^{(k)}NOT}_{1\cdots k,k+1}$} 
\label{app1}

Here, we first claim that, with a series of repetition codes, $\overline{C^{(k)}NOT}_{1\cdots k,k+1}$ can be implemented transversally. That is,
\begin{equation}
\overline{C^{(k)}NOT}_{1\cdots k,k+1} = \prod_{j=1}^N C^{(k)}NOT_{(j,1)\cdots(j,k),(j,k+1)} \;, 
\end{equation}
where $N$ is the size of each repetition code. By expanding out the matrix exponential
\begin{equation}
C^{(j)}NOT_{(j,1)\cdots(j,k),(j,k+1)}=\exp{\left[-\mathrm{i} \frac{\pi}{2^{k+1}}\left(\prod_{m=1}^k (1-Z_{j,m})\right) (1-X_{j,k+1})\right]} \;,
\end{equation}
a collection of spin-spin interactions implementing $\overline{C^{(k)}NOT}_{1\cdots k,k+1}$ can be obtained, which include Eqs.~(2) and (5) in the main text as special cases of logical $CNOT$ and $CCNOT$ gates respectively. In the following, we will prove the above claim by induction.

We start by proving that logical $CNOT$ gate can be implemented transversally. To this end, consider two repetition codes, each comprising $N$ qubits, with logical operators $\overline{Z}_A=Z_{1,A}$, $\overline{Z}_B=Z_{1,B}$, $\overline{X}_A=\prod_{j=1}^N X_{j,A}$, and $\overline{X}_B=\prod_{j=1}^N X_{j,B}$. Further, let $P_0=\prod_{j=1}^{N-1}\prod_{s=A,B}\frac{(1+Z_{j,s}Z_{j,s})}{2}$ be a projector onto the logical subspace with $Z_{j,s}Z_{j,s}=+1$. By noting that $(1+sZ_A)(1+s'Z_A)=2\delta_{s,s'}(1+sZ_A)$, $P_0Z_{j,A}P_0=P_0Z_{1,A}P_0$, and $CNOT_{(j,A),(j,B)}\propto 1+Z_{j,A}+(1-Z_{j,A})X_{j,B}$, it is straightforward to verify that the logical $CNOT$ gate between the two codes can be written as a product of physical $CNOT$ gates between their qubit constituents, i.e., $P_0\prod_{j=1}^N CNOT_{(j,A),(j,B)} P_0=P_0 \overline{CNOT}_{A,B} P_0$.

To prove the induction step, we note that $\overline{C^{(k)}NOT}_{1\cdots k,k+1}$ can be written recursively as
\begin{equation}
P_0\overline{C^{(k)}NOT}_{1\cdots k,k+1}P_0 \propto P_0(1+\overline{Z}_1)P_0 +P_0(1-\overline{Z}_1)\overline{C^{(k-1)}NOT}_{2\cdots k,k+1}P_0 \;.
\end{equation}
Since we assume that $\overline{C^{(k-1)}NOT}_{2\cdots k,k+1}$ can be implemented transversally, it can be written as $\prod_{j=1}^{N}C^{(k-1)}NOT_{(j,2)\cdots (j,k),k+1}$. By again using the identity $(1+sZ_A)(1+s'Z_A)=2\delta_{s,s'}(1+sZ_A)$, as well as $P_0\overline{Z}_{1}P_0=P_0Z_{j,1}P_0$, the proof immediately follows.

\section{Section D: Local implementation of $\overline{C^{(j)}NOT}_{A_1\cdots A_j,A_{j+1}}$} 
\label{app3}

We first note that the transversal implementation of $\overline{C^{(j)}NOT}_{A_1\cdots A_j,A_{j+1}}$ involves the many implementations of 
\begin{equation}
C^{(j)}NOT_{Z_{\ell,A_1}\cdots Z_{\ell,A_j},X_{\ell,A_{j+1}}}=\exp\left(-\mathrm{i}\frac{\pi}{2^{j+1}}\left(\prod_{k=1}^j (1-Z_{\ell,A_k})\right)(1-X_{\ell,A_{j+1}})\right)\;. 
\end{equation}
By expanding out the exponentials on the right hand side, we identify the presence of many nonlocal interactions of the form $\mathcal{I}_1 = Z_{\ell,A_{m}}Z_{\ell,A_{m+1}}\cdots Z_{\ell,A_{j}}X_{\ell,A_{j+1}}$, $\mathcal{I}_2 = Z_{\ell,A_{m}}X_{\ell,A_{j+1}}$, $\mathcal{I}_3 = Z_{\ell,A_{m}} Z_{\ell,A_{m+h}}$, and any products of these three types of interactions. In the following, we will present a scheme for breaking down $\mathcal{I}_1$, $\mathcal{I}_2$, and $\mathcal{I}_3$, into a series of local (nearest-neighbor) weight-two interactions. To this end, we note the following important identity
\begin{equation}
e^{\mathrm{i} \theta P_1} P_2 e^{-\mathrm{i} \theta P_1} = \cos(2\theta) P_1 +\mathrm{i} \sin(2\theta) P_1 P_2 \;, \label{iden}
\end{equation}
where $P_1$ and $P_2$ are anticommuting Pauli operators.

Consider an interaction of the form $\mathcal{I}_1 = Z_{\ell,A_{m}}Z_{\ell,A_{m+1}}\cdots Z_{\ell,A_{j}}X_{\ell,A_{j+1}}$. By starting with $\mathcal{I}_1^{(0)}=Z_{\ell,A_{m}}X_{\ell,A_{m+1}}$, Eq.~(\ref{iden}) implies
\begin{equation}
\mathcal{I}_1^{(1)}=e^{\mathrm{i} \frac{\pi}{4} Y_{\ell,A_{m+1}} X_{\ell,A_{m+2}}} \mathcal{I}_1^{(0)} e^{-\mathrm{i} \frac{\pi}{4} Y_{\ell,A_{m+1}} X_{\ell,A_{m+2}}} = Z_{\ell,A_{m}} Z_{\ell,A_{m+1}}X_{\ell,A_{m+2}} \;.
\end{equation}
In a similar fashion,
\begin{equation}
\mathcal{I}_1^{(2)}=e^{\mathrm{i} \frac{\pi}{4} Y_{\ell,A_{m+2}} X_{\ell,A_{m+3}}} \mathcal{I}_1^{(1)} e^{-\mathrm{i} \frac{\pi}{4} Y_{\ell,A_{m+2}} X_{\ell,A_{m+3}}} = Z_{\ell,A_{m}} Z_{\ell,A_{m+1}}Z_{\ell,A_{m+2}}X_{\ell,A_{m+3}} \;.
\end{equation}
It thus follows that $\mathcal{I}_1$ can be obtained by repeated conjugation of $\mathcal{I}_1^{(0)}$ with operators of the form $e^{\mathrm{i} \frac{\pi}{4} Y_{\ell,A_{m+k}} X_{\ell,A_{m+k+1}}}$. In particular,
\begin{equation}
e^{-\mathrm{i} \theta \mathcal{I}_1} =\left(\prod_{k=1}^{j-m} e^{\mathrm{i} \frac{\pi}{4} Y_{\ell,A_{m+k}} X_{\ell,A_{m+k+1}}} \right) e^{-\mathrm{i} \theta Z_{\ell,A_{m}}X_{\ell,A_{m+1}}} \left(\prod_{k=1}^{j-m} e^{\mathrm{i} \frac{\pi}{4} Y_{\ell,A_{m+k}} X_{\ell,A_{m+k+1}}} \right)^\dagger \;.
\end{equation}

Consider next an interaction of the form $\mathcal{I}_2=Z_{\ell,A_m}Z_{\ell,A_{m+k}}$. By starting with $\mathcal{I}_2^{(0)}=Z_{\ell,A_{m}}X_{\ell,A_{m+1}}$, we again utilize Eq.~(\ref{iden}) to grow $\mathcal{I}_2^{(0)}$ into a weight-three interaction,
\begin{equation}
\mathcal{I}_2^{(1)}=e^{\mathrm{i} \frac{\pi}{4} Y_{\ell,A_{m+1}} Y_{\ell,A_{m+2}}} \mathcal{I}_2^{(0)} e^{-\mathrm{i} \frac{\pi}{4} Y_{\ell,A_{m+1}} Y_{\ell,A_{m+2}}} = Z_{\ell,A_{m}} Z_{\ell,A_{m+1}}Y_{\ell,A_{m+2}} \;.
\end{equation} 
We then remove $Z_{\ell,A_{m+1}}$, while at the same time transforming $Y_{\ell,A_{m+2}}\rightarrow X_{\ell,A_{m+2}}$ via 
\begin{equation}
\mathcal{I}_2^{(2)}=e^{\mathrm{i} \frac{\pi}{4} Z_{\ell,A_{m+1}} Z_{\ell,A_{m+2}}} \mathcal{I}_2^{(1)} e^{-\mathrm{i} \frac{\pi}{4} Z_{\ell,A_{m+1}} Z_{\ell,A_{m+2}}} = Z_{\ell,A_{m}} X_{\ell,A_{m+2}} \;.
\end{equation}
Through repeated conjugation with the two types of operators above, one obtains
\begin{eqnarray}
e^{-\mathrm{i} \theta \mathcal{I}_2} &=&\left(\prod_{k=1}^{j-m} e^{\mathrm{i} \frac{\pi}{4} Z_{\ell,A_{m+k}} Z_{\ell,A_{m+k+1}}} e^{\mathrm{i} \frac{\pi}{4} Y_{\ell,A_{m+k}} Y_{\ell,A_{m+k+1}}} \right) e^{-\mathrm{i} \theta Z_{\ell,A_{m}}X_{\ell,A_{m+1}}} \nonumber \\
&& \times \left(\prod_{k=1}^{j-m} e^{\mathrm{i} \frac{\pi}{4} Z_{\ell,A_{m+k}} Z_{\ell,A_{m+k+1}}} e^{\mathrm{i} \frac{\pi}{4} Y_{\ell,A_{m+k}} Y_{\ell,A_{m+k+1}}}\right)^\dagger \;.
\end{eqnarray}
Finally, $\mathcal{I}_3$ can be implemented in a similar manner as $\mathcal{I}_2$, with the main difference being the use of ZX rather than ZZ operator in the second conjugation steps, i.e., 
\begin{eqnarray}
e^{-\mathrm{i} \theta \mathcal{I}_3} &=&\left(\prod_{k=1}^{h-1} e^{-\mathrm{i} \frac{\pi}{4} Z_{\ell,A_{m+k}} X_{\ell,A_{m+k+1}}} e^{\mathrm{i} \frac{\pi}{4} Y_{\ell,A_{m+k}} Y_{\ell,A_{m+k+1}}} \right) e^{\mathrm{i} \theta Z_{\ell,A_{m}}X_{\ell,A_{m+1}}} \nonumber \\
&& \times \left(\prod_{k=1}^{j-m} e^{-\mathrm{i} \frac{\pi}{4} Z_{\ell,A_{m+k}} X_{\ell,A_{m+k+1}}} e^{\mathrm{i} \frac{\pi}{4} Y_{\ell,A_{m+k}} Y_{\ell,A_{m+k+1}}}\right)^\dagger \;.
\end{eqnarray}
It should be noted that the above expansions are by no means optimal. Depending on the exact set of interactions to simulate, some products of exponentials can be simplified, e.g., by collecting commuting exponentials together and combining them as a single exponential.

As an explicit example, we apply the above procedure for locally implementing the $\overline{CCNOT}_{12,3}$ gate. According to Eq.~(5) in the main text, its transversal implementation requires the presence of two long-range interactions, i.e., $Z_{j,1}Z_{j,2}X_{j,3}$ and $Z_{j,1}X_{j,3}$. Following the above scheme for locally implementing $\mathcal{I}_1$ and $\mathcal{I}_2$ types of interactions, we obtain
\begin{eqnarray}
\overline{CCNOT}_{12,3} &=& e^{-\mathrm{i} \sum_{j=1}^N \frac{\pi}{8} \left[Z_{j,1}Z_{j,2}+Z_{j,2}X_{j,3}-Z_{j,1}-Z_{j,2}-X_{j,3}\right]} \times e^{\mathrm{i} \sum_{j=1}^N \frac{\pi}{4} Y_{j,2}X_{j,3} } \nonumber \\
&\times& e^{\mathrm{i} \sum_{j=1}^N \frac{\pi}{8} Z_{j,1}X_{j,2} } \times e^{-\mathrm{i} \sum_{j=1}^N \frac{\pi}{4} Y_{j,2}X_{j,3}} \times e^{\mathrm{i} \sum_{j=1}^N \frac{\pi}{4} (Z_{j,2}Z_{j,3}+Y_{j,2}Y_{j,3})} \nonumber  \\
&\times& e^{-\mathrm{i} \sum_{j=1}^N \frac{\pi}{8} Z_{j,1}X_{j,2} } \times e^{-\mathrm{i} \sum_{j=1}^N \frac{\pi}{4} (Z_{j,2}Z_{j,3}+Y_{j,2}Y_{j,3})} \;.
\end{eqnarray}    

\section{Section E: Quantum circuit simulating $4T$-periodic DTC in Sycamore processor}
\label{app4}

The two-qubit gate operations simulating the various spin-spin interactions required for large period DTCs may not be inherently native within the Sycamore device. However, it is possible to build such two-qubit gate operations using appropriate combinations of single-qubit rotations and the $iSWAP_{A,B}=e^{-\mathrm{i}\frac{\pi}{4}(X_AX_B+Y_AY_B)}$ gate, both of which are Sycamore's native gates \cite{Sycamore,DTCqs2}. More explicitly, a two-qubit $ZX$ ($ZY$) rotation is obtained by conjugating a single $Y$-($X$-)rotation with the $iSWAP$ gate, i.e.,
\begin{eqnarray}
\exp\left(-\mathrm{i} \theta Z_{1}X_{2}\right) &=& iSWAP_{1,2} \exp\left(-\mathrm{i} \theta Y_1\right) iSWAP_{1,2}^{(-1)} \;, \nonumber \\
\exp\left(-\mathrm{i} \theta Z_{1}Y_{2}\right) &=& iSWAP_{1,2}^{(-1)} \exp\left(-\mathrm{i} \theta X_1\right) iSWAP_{1,2} \;. \label{ZX}
\end{eqnarray} 
Other two-qubit gates, such as the $ZZ$ rotation necessary for the implementation of repetition code Hamiltonians, can then be obtained by further conjugating Eq.~(\ref{ZX}) with appropriate $\pi/4$ single qubit rotation.

\begin{center} 
	\begin{figure}
		\includegraphics[scale=0.8]{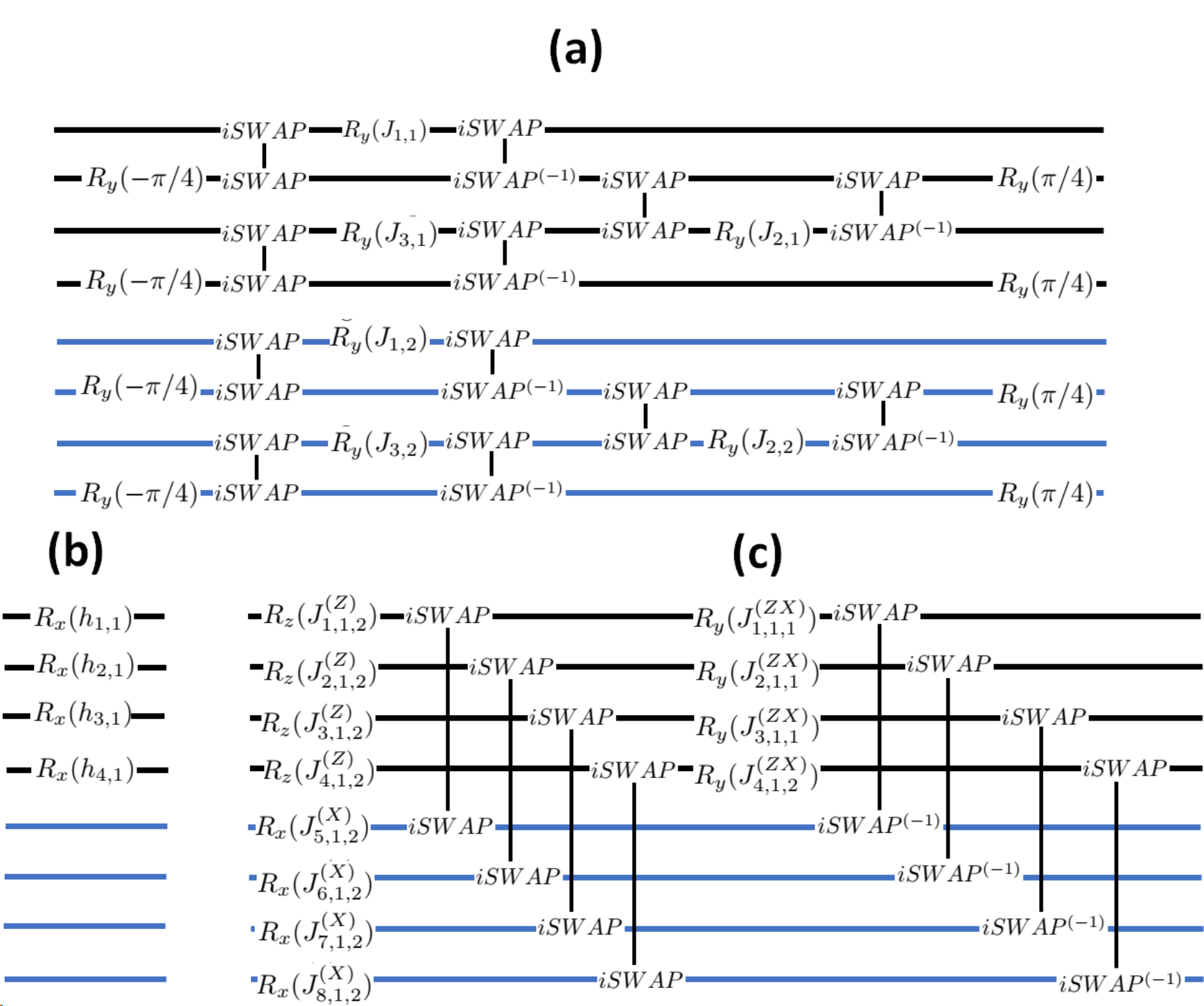}
		\caption{Quantum circuit schematics of (a) $e^{-\mathrm{i} [H_{\rm rep,1}+H_{\rm rep,2}]}$, (b) $\overline{X}_1$, and (c) $\overline{CNOT}_{1,2}$ for two copies of size-four quantum repetition codes using a combination of $iSWAP$ and single-qubit rotations. The first and second repetition codes are marked with black and blue colours respectively. Note that a series of inter-code $iSWAP$s appearing in (c) can be applied parallelly.}
		\label{cirsyc}
	\end{figure}
\end{center}

With the above in mind, a possible quantum circuit simulation of the proposed $4T$-periodic DTC is summarized in Fig.~\ref{cirsyc}(a-c), which is directly obtained from the explicit expression of $U_T^{(4)}$ (Eq. (1) in the main text) by breaking down all weight-two interactions into $iSWAP$s and single qubit rotations, i.e., $R_S(\theta)=e^{-\mathrm{i} \theta S}$ for $S=X,Y,Z$, according to Eq.~(\ref{ZX}). In this case, all the necessary intentional disorder can be programmed with respect to the single qubit rotations $R_y(J_{j,s})$, $R_x(h_{j,s})$, $R_z(J_{j,1,2}^{(Z)})$, $R_x(J_{j,1,2}^{(X)})$, and $R_y(J_{j,1,2}^{(ZX)})$. Moreover, while the circuit naturally starts in the $|0\cdots 0\rangle$ state, we may apply appropriate single qubit rotations to explore the robustness of the simulated DTCs under different choices of initial states. A larger version of such a circuit, which represents two copies of size-eight quantum repetition codes, was used to produce Fig.~4 in the main text. Finally, it is worth mentioning that in the actual experiments, further simplification of the circuit is expected, e.g., by redefinition of the initial state and simplification of some single qubit rotations.   

\end{document}